# Structure and rheology of multi-chain amphiphilic block copolymers under shear in dilute solutions

Ehsan Kamali Ahangar, Dominic Robe, Elnaz Hajizadeh[*]

Soft Matter Informatics Research Group, Department of Mechanical Engineering, The University of Melbourne, Victoria 3010, Australia

[*]Corresponding author: ellie.hajizadeh@unimelb.edu.au

**Abstract**

This study presents a computational investigation of self-assembly and rheological behaviour of multichain amphiphilic block copolymers under varying chain length, architecture, composition, and shear rate. Using Brownian dynamics (BD) simulations, we systematically examined bead-spring model multi-chain diblock and triblock copolymers with chain lengths of 12-48 beads, hydrophobic fractions (*f*) ranging from 0 to 1.0, and shear rates spanning $\dot{\gamma}$=0-0.1 $ns^{-1}$. In the dilute regime, results demonstrate that triblock copolymers form extensive 3D networks with bridging architectures through hydrophobic end blocks, achieving solution viscosities up to half an order of magnitude higher than diblock systems, with superior structural integrity under weak shear. At $\dot{\gamma}$=0.003-0.01 $ns^{-1}$, both chain architectures show increased gyration radius of individual chains within each micelle and decreased cluster counts, indicating aggregation of clusters prior to breakdown at higher shear rates. Shape anisotropy analysis reveals that triblocks develop highly elongated prolate structures ($L_1/L_3 \approx 11$) at high shear rates, while diblocks form more discrete micellar assemblies ($L_1/L_3 \approx 7.5$). Chain length analysis shows systematic increases in radius of gyration, with triblocks exhibiting an increase in cluster count, indicative of network percolation. Rheologically, triblock systems maintain lower crossover frequencies with increasing hydrophobic fraction, reflecting slower network relaxation versus diblocks. The terminal relaxation time of triblock copolymer systems increases with hydrophobic fraction due to double-ended hydrophobic bridging, while diblocks maintain stable values. These findings provide fundamental insights for the rational design of polymer-based drug carriers through architectural selection and flow conditions.

## 1. Introduction

Amphiphilic block copolymers have emerged as a versatile class of polymers with applications spanning enhanced oil recovery, drug delivery, water treatment, gene therapy, and advanced materials engineering [1], [2], [3], [4], [5]. These macromolecules, consisting of chemically distinct blocks with contrasting affinities toward solvents, possess unique properties that enable them to address critical challenges in nanomedicine, biotechnology, and soft matter research [6]. The self-assembly behaviour of amphiphilic block copolymers in selective solvents is governed by the delicate interplay between interfacial tension at the core-corona boundary, excluded volume interactions in the corona, and chain stretching penalties in both domains. When dissolved in selective solvents, these copolymers spontaneously self-assemble into a rich variety of nanostructures including spherical micelles, cylindrical or wormlike micelles, vesicles (polymersomes), lamellae (bilayers), and more complex multicompartment architectures [7]. The formation of these diverse morphologies depends on the relative volumes of hydrophobic and hydrophilic blocks, interfacial area per chain, and core chain length [8]. Early theoretical frameworks established by Birshtein and Zhulina [9] in their scaling theory of supermolecular structures in block copolymer-solvent systems provided fundamental insights into how equilibrium micellar parameters depend on copolymer composition, solvent strength, and polymer concentration. Building upon this foundation, Zhulina and Borisov [10] presented a comprehensive review of block polymer micelle theory, delineating the distinctions between star-like and crew-cut regimes based on the relative length of corona and core blocks. The star-like regime occurs when the corona block is significantly longer than the core block, resulting in extended, brush-like coronas with relatively small cores. In contrast, the crew-cut regime is characterized by a short corona block relative to the core block, producing micelles with large, well-defined hydrophobic cores and relatively thin coronas. The current study focuses on the crew-cut regime (the most commonly observed) as well as the star-like regime, in which the identified potential morphologies of block copolymers exhibit these characteristics. They also established power-law scaling relationships that connect molecular architecture to micellar dimensions and aggregation numbers in both regimes, providing predictive design rules that have been validated across numerous experimental systems.

Experimental characterization techniques have provided critical validation for theoretical predictions while revealing phenomena not easily accessible through theory alone. Cryogenic transmission electron microscopy (cryo-TEM) has emerged as the gold standard for direct visualization of self-assembled block copolymer structures in their native hydrated state. Won et al. [8] employed cryo-TEM to investigate micellar polymorphism of poly(ethylene oxide)-based block copolymers, determining boundaries for shape transitions from bilayers to cylinders to spheres with increasing PEO composition through direct imaging of vitreous hydrated specimens. Their analysis enabled the determination of packing properties including interfacial area per chain and degree of chain stretching in the hydrophobic core. Solvent-induced morphological transitions have been systematically investigated through combined experimental techniques. Campos-Villalobos et al. [11] demonstrated that controlled variation of solvent composition induces reversible transitions between spheres, rods, and vesicles in methacrylate-based block copolymer aggregates, establishing that morphology can be kinetically trapped during dialysis processes. Combined experimental techniques were employed by Sabadini et al. [12] to investigate ethoxylated complex coacervate core micelles (C3Ms) using dynamic light scattering (DLS), SAXS, and cryo-TEM, revealing temperature-induced transitions from spherical to elongated aggregates contingent on poly(ethylene oxide) block length, with behaviour paralleling classical ethoxylated surfactant aggregates. Multiple asymmetric amphiphilic diblock copolymer systems have been characterized through physico-chemical investigations, revealing how subtle variations in block length ratios and chemical composition produce dramatically different self-assembled morphologies in solution [13]. These morphological transitions have important implications for the design and synthesis of ordered mesoporous materials, where block copolymer micelles serve as structure-directing agents [14]. Computational studies encompass a range of techniques, such as Brownian dynamics, molecular dynamics [56], [57], [58], Monte Carlo simulations, and dissipative particle dynamics, providing mechanistic insights into assembly pathways and structure-property relationships that complement experimental observations. Brownian dynamics simulations provide complementary insights into self-assembly mechanisms by explicitly accounting for stochastic thermal fluctuations while maintaining computational tractability for large systems. Hafezi and Sharif [15] performed systematic BD simulations of amphiphilic diblock copolymers with varying hydrophobic tail lengths, demonstrating that critical micelle concentration (CMC) exhibits logarithmic proportionality to tail length, while aggregation number increases with hydrophobic block length and decreases with hydrophilic block length according to power-law scaling relationships. Importantly, their comparison with mixed block copolymer systems (comicelles) revealed that equilibrium structural and dynamic parameters depend similarly on the actual chemical length of the hydrophobic block (intrinsic length) versus the effective hydrophobic length in mixed systems where different copolymer types co-assemble (apparent length, determined by the average hydrophobicity experienced in the mixed core), a finding with direct implications for designing polymeric micelle carriers with controlled stability in the bloodstream, where concentrations fall below CMC. Specialized Brownian dynamics simulations of associating diblock copolymers have been developed to capture the reversible binding dynamics of end-functionalized chains, revealing how sticker association lifetimes influence network structure and relaxation behaviour [16].

For multiblock architectures, Zhang et al. [17] investigated self-assembly structures of amphiphilic multiblock copolymers in dilute solution as functions of solvent quality and backbone stiffness, identifying diverse morphologies including single-flower micelles, multi-flower micelles, and single or multi-bridge structures. Their phase diagrams demonstrated that highly hydrophobic components favour flower micelle formation while semi-flexible chains preferentially form bridge structures, with chain length increases driving single-to-multi-flower transitions. The influence of architecture on unimolecular micelle formation was examined by Hao et al. [18] for amphiphilic comb-like copolymers in selective solvents, revealing the existence of an optimal total chain length for spherical micelle stability beyond which the spherical morphology transforms into cylindrical micelles. Their analysis of radius of gyration and stretching factor showed exponential relationships with power exponents decreasing as hydrophilic side chain length increased. Molecular dynamics (MD) simulations with explicit and implicit solvent models have enabled detailed investigations of morphological diversity and assembly pathways. Liu and Sureshkumar [19] employed coarse-grained MD with explicit water-mediated hydrophilic/hydrophobic interactions to track the spatiotemporal evolution of diblock copolymer aggregation from initially homogeneous solutions, successfully capturing experimentally observed morphological diversity including spherical micelles, vesicles (polymersomes), lamellae (bilayers), linear wormlike micelles, and toroidal structures. Their comprehensive analysis quantified aggregation numbers, packing parameters, radial distribution functions, and energetic/entropic measures such as solvent accessible surface area (SASA) and probability distribution functions of segmental chain stretch. Building upon this foundation, Liu et al. [20] investigated vesicle morphogenesis in amphiphilic BAB triblock copolymer solutions, identifying the vesiculation pathway as progressing through interconnected networks of copolymer aggregates to cage structures and finally to mature vesicles, with the assembly process analysed using information entropy metrics. Benchmark coarse-grained MD methodology was established by Srinivas et al. [21], demonstrating self-assembly and property prediction capabilities for diblock copolymers. Multiscale modelling approaches were pioneered by Bedrov et al. [22], who integrated quantum chemistry calculations, atomistic explicit solvent MD, and coarse-grained implicit solvent simulations to predict physical properties of poly(ethylene oxide)-poly(propylene oxide)-poly(ethylene oxide) Pluronic triblock copolymer micelles in aqueous solution with systematic elimination of computationally expensive degrees of freedom. The role of hydrodynamic interactions in collapse dynamics was investigated by Pham et al. [23] using Brownian dynamics for copolymers subjected to sudden solvent quality quenches from good to poor, demonstrating that chain sequence strongly influences collapse kinetics, compactness, and energy of the final globular state through pathways involving rapid cluster formation followed by coalescence and rearrangement.

Flow-induced structural transformations have been directly observed through rapid vitrification techniques combined with microscopy. Understanding stress-structure relationships is essential for the processing and application of reversible associating polymer networks. Recent experimental investigations have revealed that under start-up shear flow, the transient stress response exhibits strong coupling to the evolution of network connectivity and micelle structure, with stress overshoot phenomena arising from the competition between flow-induced chain stretching and dissociation of associative junctions [24,45]. Rheological investigations have demonstrated that the nonlinear response of supramolecular polymer networks formed by associative telechelic chains depends sensitively on both shear and extensional flow character, with different chain deformation mechanisms dominating in each flow geometry [25]. Rheological behaviour and shear flow effects on block copolymer solutions have been investigated through specialized simulation techniques incorporating hydrodynamic interactions and non-equilibrium conditions. Sugimura and Ohno [26] developed Monte Carlo Brownian Dynamics simulations incorporating the bond fluctuation model for chain flexibility and Kramers potential for shear forces, applying this methodology to water/oil/ABA triblock copolymer ternary systems. Their analysis revealed that shear viscosity and shear stress both increase exponentially with polymer concentration, with micelle patterns transforming from thermal equilibrium states to steady-state configurations under flow [47]. Simulations specifically targeting shear-induced morphological transitions in block copolymers have demonstrated that applied flow fields can drive transitions between lamellar, cylindrical, and spherical morphologies, with the direction and magnitude of structural changes depending on flow strength, copolymer composition, and initial morphology [27]. Dissipative particle dynamics (DPD) has emerged as a particularly powerful tool for studying flow effects due to its inherent incorporation of hydrodynamic interactions and computational efficiency at mesoscales. Nonequilibrium DPD simulations of rod-coil diblock copolymers in solution have revealed that shear flow induces alignment of cylindrical micelles along the flow direction, with the degree of orientation increasing with shear rate until a critical value where micelle fragmentation occurs [28]. Advanced simulation methodologies for block copolymer rheology were implemented by Schneider and Müller [29] using GPU-accelerated HOOMD-blue software with reverse non-equilibrium molecular dynamics [60] simulation to enable oscillatory shear flow studies and nonlinear rheology investigations. Their systematic analysis of lamellae-forming diblock copolymer melts revealed that perpendicularly oriented lamellae exhibit rheological properties similar to unstructured homopolymer melts, while parallel and transverse orientations show similar dynamic-mechanical behaviour in the linear regime characterized by high dissipation, with molecular configurations providing further insights into single-chain dynamics.

Despite substantial progress in understanding block copolymer self-assembly under both quiescent and flow conditions, the full spectrum of flow-induced structural transformations across micellar systems, from isolated micelles to interconnected networks, remains poorly understood. A critical and largely overlooked question is how chain length and hydrophobic fraction govern the structural and rheological response when subjected to flow [59]. To the best of the authors' knowledge, this is the first computational study to systematically address this question for amphiphilic diblock and triblock copolymers in dilute solution under shear flow, revealing non-trivial and architecture-dependent behaviours with direct implications for the rational design of polymer-based drug carriers. The paper is organized as follows. In Section 2, the problem is stated, input parameters are defined, and the simulation approach, including the model and boundary conditions, is described. In Section 3, comprehensive results and discussion are presented, covering chain conformation and morphology, gyration radius, cluster count, solution viscosity, viscoelasticity (moduli G′ and G″), and terminal relaxation time, followed by a summary of findings in Section 4.

## 2. Methodology

### 2.1. Problem Statement and Copolymer Coarse-Grained Model

In Figure 1, the potential morphologies of block copolymers observed in this study are shown, including spherical micelles, cigar-shaped micelles, short cylinder-like micelles, and worm-like micelles. Figure 2 demonstrates initial configurations of studied systems that serve as starting points to investigate the evolution of nanostructure from disordered to ordered across the design space. Figure 2a shows the initial configurations for two types of molecular architectures: diblock copolymers ($N_{block} = 2$) and triblock copolymers ($N_{block} = 3$), with varying chain lengths of $N = 12, 24, 36$, and $48$ beads per chain. The number of polymer chains ($N_{chain}$) is fixed at 120 to examine the self-assembly behaviour of block copolymers in the dilute regime, while the hydrophobic fraction is set to $f = 0, 0.25, 0.5, 0.75$ or $1$ in Figure 2b. External flow conditions include both quiescent condition and applied shear rates at $\dot\gamma = 0, 0.003, 0.01, 0.03$, and $0.1\ ns^{-1}$. Figure 2c provides a schematic illustration of the block architecture for diblock and triblock copolymers with $N = 12$ beads, demonstrating how the composition parameter $f$ controls the distribution of hydrophobic (A) and hydrophilic (B) segments within each chain. For diblock copolymers ($N_{block} = 2$), the architecture consists of two distinct blocks: ($B_{12}$) for $f = 0$ (purely hydrophilic), ($A_3$–$B_9$) for $f = 0.25$. In contrast, triblock copolymers ($N_{block} = 3$) exhibit a more complex architecture with three sequential blocks: ($A_3$–$B_6$–$A_3$) for $f = 0.5$, representing a symmetric ABA configuration where hydrophobic end blocks flank a central hydrophilic block, and ($A_{12}$) for $f = 1$ (purely hydrophobic). In all cases shown in the results and discussion section, the volume fraction ($\phi$) in the dilute regime is fixed at 0.07193.

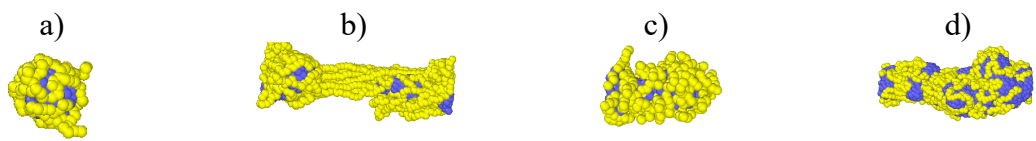

Figure 1: Potential morphologies of multi-chain amphiphilic block copolymers observed in the current study a) spherical micelle; b) cigar-shaped micelle; c) short cylinder-like micelle; and d) worm-like micelle

$N_{chain} = 120$ and $f = 0.5$

| $N = 12$ | $N = 24$ | $N = 36$ | $N = 48$ |
|---|---|---|---|
| $N_{block} = 2$ | $N_{block} = 2$ | $N_{block} = 3$ | $N_{block} = 3$ |

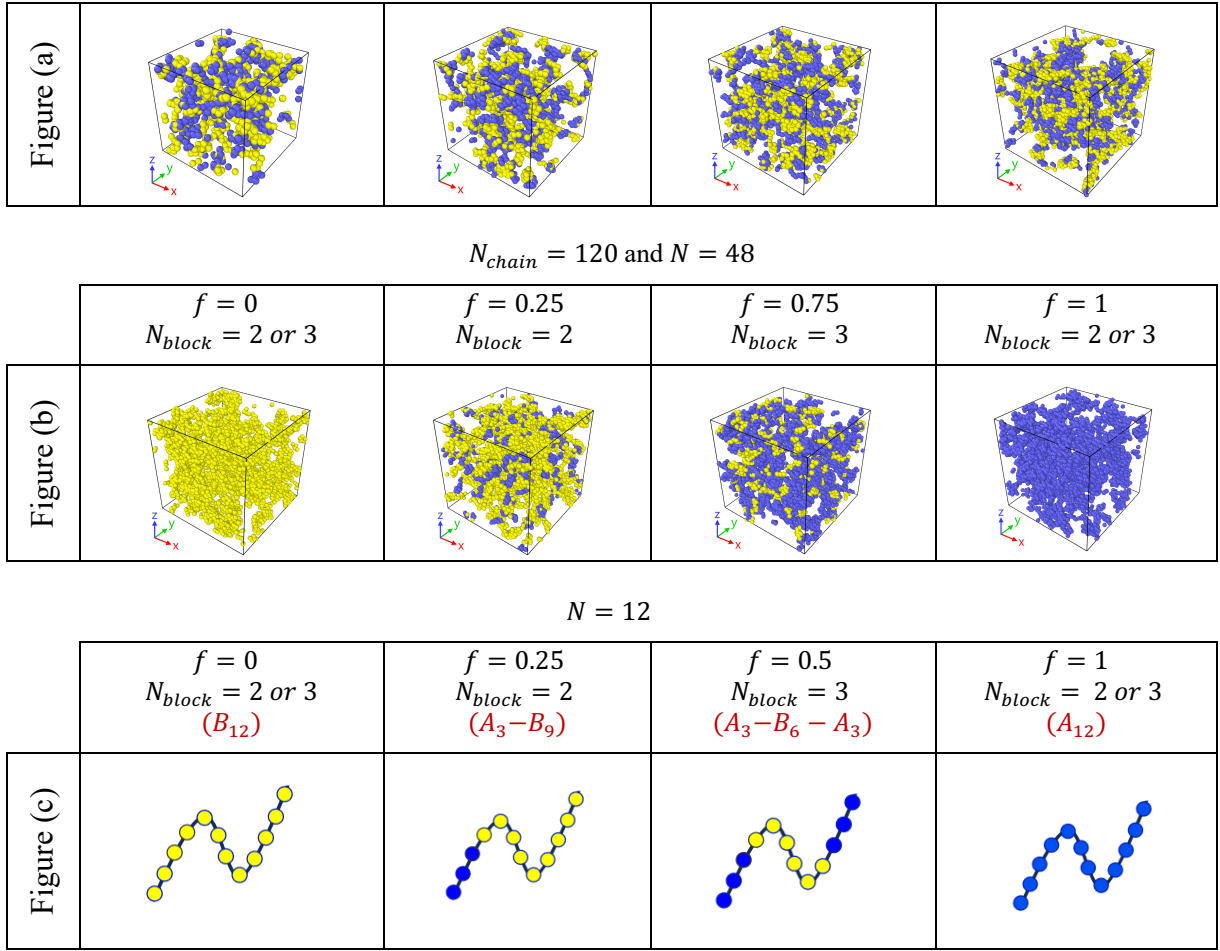

Figure 2: Initial configurations of multi-chain diblock and triblock copolymers studied at a) various chain lengths, and fixed fraction parameter f=0.5; b) different composition parameters and fixed chain length N=48; c) various hydrophobic fraction parameters affecting a schematic illustration of the block architecture with 12 beads (hydrophobic and hydrophilic beads are shown in blue and yellow colours, respectively)

The intermolecular interactions are described by Lennard-Jones (LJ) potentials as follows [18]:

$$U_{LJ}(r_{ij}) = 4\epsilon \left[ \left(\frac{\sigma}{r_{ij}}\right)^{12} - \left(\frac{\sigma}{r_{ij}}\right)^{6} \right] \quad (1)$$

where the energy and length are respectively given by $\epsilon$ and $\sigma$ (with different interaction strengths and cutoff distances depending on the bead types). For hydrophobic-hydrophobic interactions (type 2-2), we use $\epsilon_{22}$ = 4.195 *agr.nm²/ns²*, σ = 1.5 *nm*, and $r_{cut}$ = 3.75 *nm*, representing strong attractive interactions that promote core-shell micelle formation. Hydrophilic-hydrophilic interactions (type 3-3) are characterized by $\epsilon_{33}$ = 0.524 *agr.nm²/ns²*, σ = 1.5 *nm*, and $r_{cut}$ = 1.68369 *nm*, reflecting weaker interactions appropriate for solvated segments. Cross-interactions between hydrophobic and hydrophilic beads (type 2-3) use $\epsilon_{23}$ = 0.786 *agr.nm²/ns²*, σ = 1.5 *nm*, and $r_{cut}$ determined by the hydrophobic bead parameters, representing intermediate interaction strength at the micelle core-corona interface.

Bonded interactions between consecutive beads along the polymer backbone are described by Finitely Extensible Nonlinear Elastic (FENE) potentials [18]:

$$U_{FENE}(r_{ij}) = -0.5\kappa_b R_m^2 \ln\left[1 - (\frac{r_{ij}}{R_m})^2\right] \tag{2}$$

where $\kappa_b$ and $r_{ij}$ denote the spring constant and the distance between two neighbouring beads. $R_m$ represents the maximum separation distance for $r_{ij}$ ($\kappa_b$=60, $R_m$=2.25nm). The LJ component uses the appropriate ε and σ values for the specific bond type. Three bond types are employed: type 1 for hydrophobic-hydrophobic bonds, type 2 for hydrophilic-hydrophilic bonds, and type 3 for hydrophobic-hydrophilic cross-bonds.

## 2.2. Brownian Dynamics Simulation

We employ the Brownian dynamics (BD) simulation technique to model the polymer behaviour in an implicit solvent environment. The BD approach treats solvent effects implicitly through friction and random force terms, making it computationally efficient for studying large-scale phenomena such as self-assembly in block copolymers. The equation of motion for each bead is governed as follows [17]:

$$r_i(t + \Delta t) = r_i(t) + v_i(t)\Delta t + \frac{\Delta t}{\gamma}(F_i^C + F_i^R) \tag{3}$$

where $r$ and $v$ are the position and velocity of the bead ($i$), respectively, $F_i^C$ represents conservative forces from bonded and non-bonded interactions on bead ($i$), $F_i^R$ is the random force satisfying the fluctuation-dissipation theorem, and γ is the friction coefficient with γ = $k_B T/D$, where $D$ is the diffusion coefficient.

The effect of the solvent is implicitly treated by the random force, which satisfies the fluctuation-dissipation theorem [17]:

$$\langle F_i^R(t) F_j^R(\acute{t}) \rangle = -6\gamma k_B T \delta_{ij} \delta(t - \acute{t}) \tag{4}$$

where the coupling between friction and random forces acts as an effective thermostat. The random forces are generated using a Gaussian random number generator to ensure proper thermal fluctuations. Integration is performed using a time step *dt* = 0.00005 *ns* with the Brownian integrator implemented in LAMMPS software, which is fed into Python-based scripts for morphology-dependent analysis. In Equation 4, $k_B$, $T$, and $\delta$ represent the Boltzmann constant, temperature, and Dirac delta function $\delta(t - \acute{t})$ ensures that random forces at different times are uncorrelated, respectively.

## 2.3. Simulation Procedure and Comparison of BD Simulation with Flory Theory

The simulation procedure is shown in Figure S1, with more details on the required computational libraries provided in Section S1. The gyration radii of the chains extracted from the BD simulation are compared with those from Flory theory (using calculated scaling exponents for both good and poor solvent conditions as derived from established theoretical and experimental data), with details provided in Section S2. The present model employs a generic coarse-grained bead-spring representation in which each bead constitutes an abstract interaction site characterising hydrophobic or hydrophilic chemical identity, and does not correspond to a mapped group of atoms from any specific polymer chemistry, as in systematic coarse-grained approaches such as MARTINI [30], Iterative Boltzmann Inversion-based models [31], and , learning-based studies [32], [33], [46], [52], [51], [50], [49], [48]. Consistent with established practice for generic coarse-grained models [34], [35], [55], [54], [53] this deliberate modelling choice enables systematic exploration of a wide architectural and compositional design space within a unified framework, with validation performed at the level of universal scaling laws and qualitative trends. The Flory scaling validation confirms correct conformational physics [36], [37], while the observed shear-thinning behaviour is in qualitative agreement with experimental observations for block copolymer micellar solutions [38], [39].

## 3. Results and Discussion

### 3.1. Chain Conformation and Morphology of Block Copolymers

In Figures 3 and 4, the integration of visual structural evolution and morphological analysis reveals the complete physical picture of how block copolymer architectures respond to shear flow, demonstrating the fundamental mechanisms controlling self-assembly and deformation in these systems. In Figure 4, $L_1$, $L_2$, and $L_3$ (in descending order) represent the principal components (eigenvalues) of the gyration tensor, and there is a direct and fundamental relationship with the radius of gyration as $Rg^2 = L_1+L_2+L_3$. The structural evolution of the triblock copolymer under increasing shear rate demonstrates the unique capabilities of bridging architectures. In Figure 3a, at equilibrium ($\dot{\gamma} = 0$), extensive three-dimensional networks span the simulation volume, with yellow hydrophilic chains connecting blue hydrophobic nodes in a percolating structure [40]. This network structure gives rise to the dramatic morphological response observed in the shape anisotropy data in Figure 4a, where the $L_1/L_3$ ratio reaches approximately 11 at low shear rates before declining at higher rates. As shear rate increases to $\dot{\gamma} = 0.003$-$0.01$ $ns^{-1}$, the networks orient and stretch along the flow direction while preserving their interconnected character. The visual snapshots show how individual triblock chains simultaneously participate in multiple assembly regions, creating mechanically robust structures that undergo significant deformation while maintaining connectivity. At higher shear rates ($\dot{\gamma} = 0.03$-$0.1$ $ns^{-1}$), progressive network fragmentation becomes evident as continuous structures break into smaller, more compact micelles [41], [42]. The morphological data show decreasing anisotropy ratios ($L_1/L_3$ declining from ~11 to ~10, $L_1/L_2$ from ~5.5 to ~4.8), reflecting the system's response to excessive stress through structural reorganization rather than continued elongation. The diblock copolymer exhibits fundamentally different structural responses throughout the shear range. At equilibrium ($\dot{\gamma} = 0$), discrete spherical micelles with clear core-shell organization dominate the system. The visual snapshots show these assemblies as distinct yellow-blue structures distributed throughout the simulation volume. Under shear, these micelles deform into elongated structures, while maintaining their discrete character, corresponding to lower shape anisotropies that remain relatively stable across high shear rates ($L_1/L_3 \approx 7.5$). The $L_1/L_2$ and $L_2/L_3$ ratios reveal distinct three-dimensional deformation patterns between architectures. Triblock systems show clear separation between eigenvalue ratios under shear, with $L_1/L_2$ reaching ~5 while $L_2/L_3$ remains ~2-3, indicating prolate (cigar-shaped) deformation where networks elongate preferentially along the flow direction while maintaining cross-sectional integrity through bridging connections. The visual snapshots at $\dot{\gamma} = 0.01$ $ns^{-1}$ and $0.03$ $ns^{-1}$ clearly show this elongated morphology oriented with the flow. Diblock copolymers exhibit more modest and balanced $L_1/L_2$ and $L_2/L_3$ values (~3-4 and ~2 respectively) at moderate shear rates, reflecting more symmetric deformation, where discrete micelles elongate under shear while preserving greater symmetry compared to triblock networks. The snapshots confirm the maintenance of more compact, less extensively deformed structures throughout the shear rate range.

In Figures 3b and 4b, the chain length dependence reveals distinct behaviours that reflect the fundamental differences between diblock and triblock architectures. In Figure 4b, for triblocks at $\dot{\gamma} = 0.01\ ns^{-1}$, the shape anisotropy ratios show smooth increasing trends with chain length: $L_1/L_3$ increases from ~7.5 at N = 12 to ~10 at N = 48, and $L_1/L_2$ rises from ~4 to ~5, while $L_2/L_3$ remains relatively stable around 2-2.5 throughout the range. The visual snapshots demonstrate the structural origin of this behaviour. In Figure 3b under the same volume fraction, at N = 12, small spherical and short cylinder-like micelles are observed; at N = 24, more extended interconnected networks begin to form; at N = 36, well-developed percolating networks with multiple interconnected domains span the simulation volume; at N = 48, the system transitions to fewer but more extensively elongated network structures. This progression reflects how longer chains enable greater bridging distances and more efficient network association, with the reduction in the number of discrete assemblies at higher molecular weights allowing the formation of larger, more cohesive networks that can achieve greater elongation along the flow direction while maintaining their characteristic interconnected network structure. Diblock systems exhibit the same trend in shape anisotropy with chain length. Notably, diblocks achieve higher $L_1/L_3$ and $L_1/L_2$ values than triblocks at equivalent chain lengths despite forming discrete micelles rather than percolating network structures. This counterintuitive result arises from the fundamental architectural constraints. Diblock micelles, lacking bridging segments, deform as independent entities achieving extreme elongation ratios without topological constraints. As chain length increases, these isolated micelles grow larger and undergo more extensive flow-induced deformation, directly increasing shape anisotropies. Conversely, triblock networks must balance elongation with maintaining connectivity across junction points, limiting maximum aspect ratios.

The shape anisotropy trends vary systematically with composition and block architecture. In Figures 3c and 4c, at f = 0 (pure hydrophilic), $L_1/L_3$ triblocks exhibit maximum elongation with ratios reaching ~38-39 for $N_{block}$=2, indicating extreme prolate deformation under flow. $L_1/L_2$ triblocks show intermediate anisotropy (~13), while $L_2/L_3$ triblocks display minimal change (~4). As f increases to 0.25, all triblock systems show sharp decreases in aspect ratios: $L_1/L_3$ drops to ~16, $L_1/L_2$ to ~7, and $L_2/L_3$ to ~3, as emerging hydrophobic domains disrupt alignment and double-ended micelles also reduce the chain extensibility of the middle hydrophilic block. At f = 0.5 and 0.75, continued micellization corresponds to further ratio reductions, with $L_1/L_3$ decreasing to ~10 and ~8 (worm-like micelle), respectively, while $L_2/L_3$ approaches ~2-3. At f = 1.0 (pure hydrophobic), all systems converge toward minimal ratios (~2-4), reflecting more compact structures, where hydrophobic aggregation dominates over flow-induced deformation. Diblocks display lower shape ratios compared to triblocks across all compositions (shown by solid lines below dashed lines), indicating that the absence of a middle block (as a linker) limits the overall chain extensibility under flow. At f = 0, both architectures exhibit identical values due to purely hydrophilic composition. Starting from f = 0.25, diblocks show significantly reduced deformability, where $L_1/L_3$ reaches ~7 versus ~16 for triblocks, with this performance gap persisting through intermediate compositions (f = 0.5-0.75) before converging at f = 1.0, where compact hydrophobic aggregation dominates.

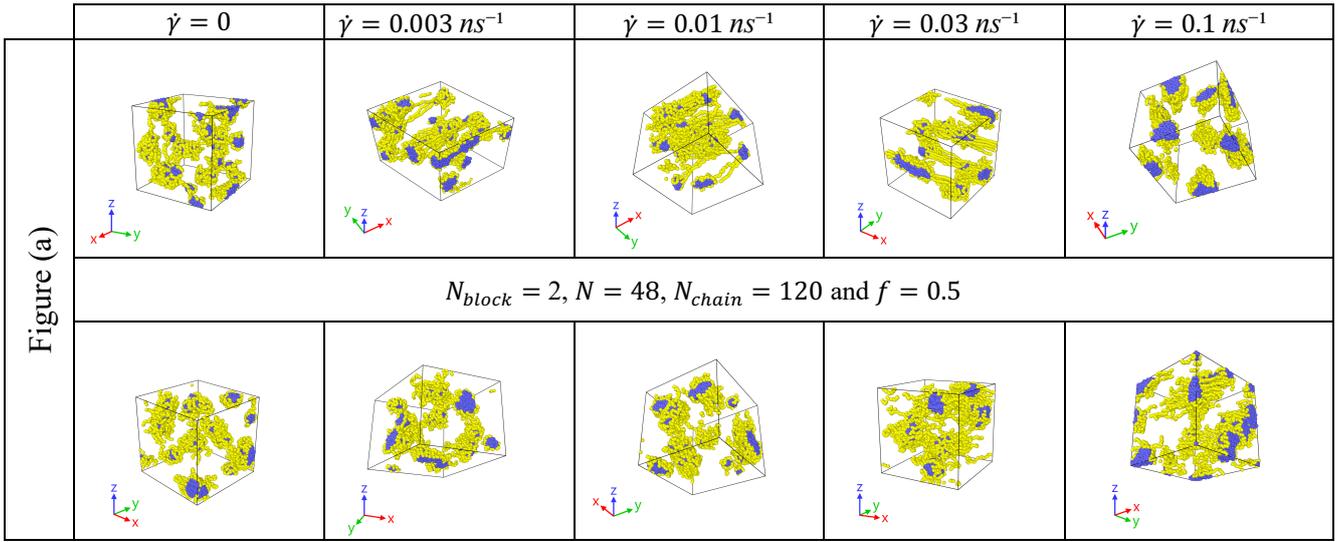

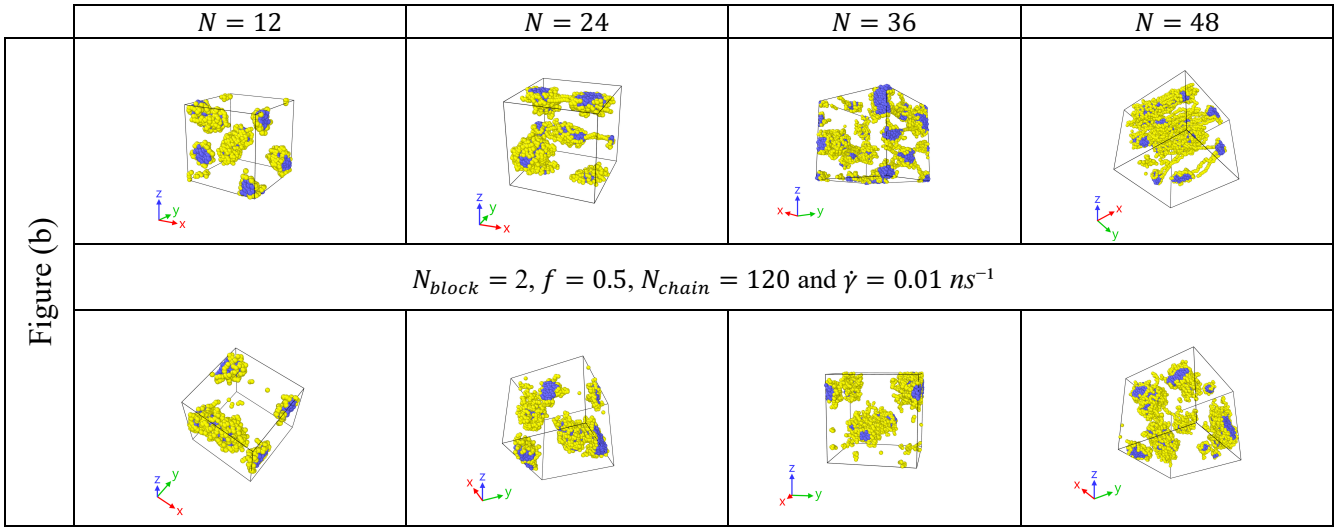

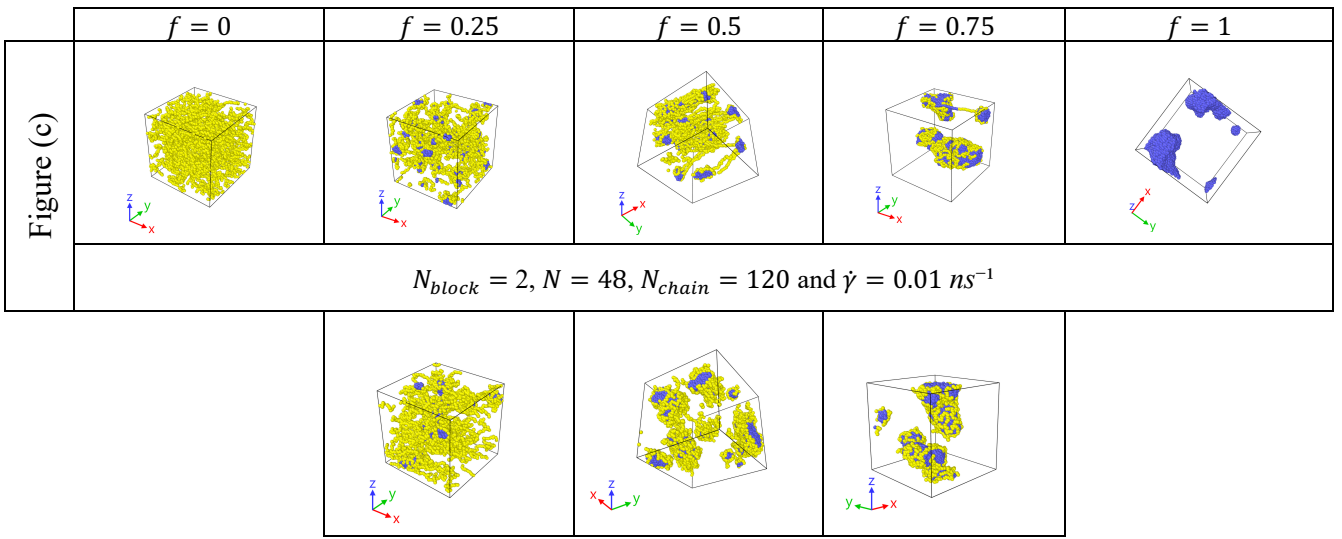

Figure 3: a) The effect of shear rate on the chain confromation for both diblock and triblock copolymers with 120 chains, chain length N= 48 beads and hydrophobic fraction of 0.5; b) The impact of chain length on chain

conformation of both multi-chain copolymers at f=0.5 and shear rate of 0.01 $ns^{-1}$; c) The effect of composition parameter on chain conformation of both multi-chain copolymer systems at N=48 and shear rate of 0.01 $ns^{-1}$

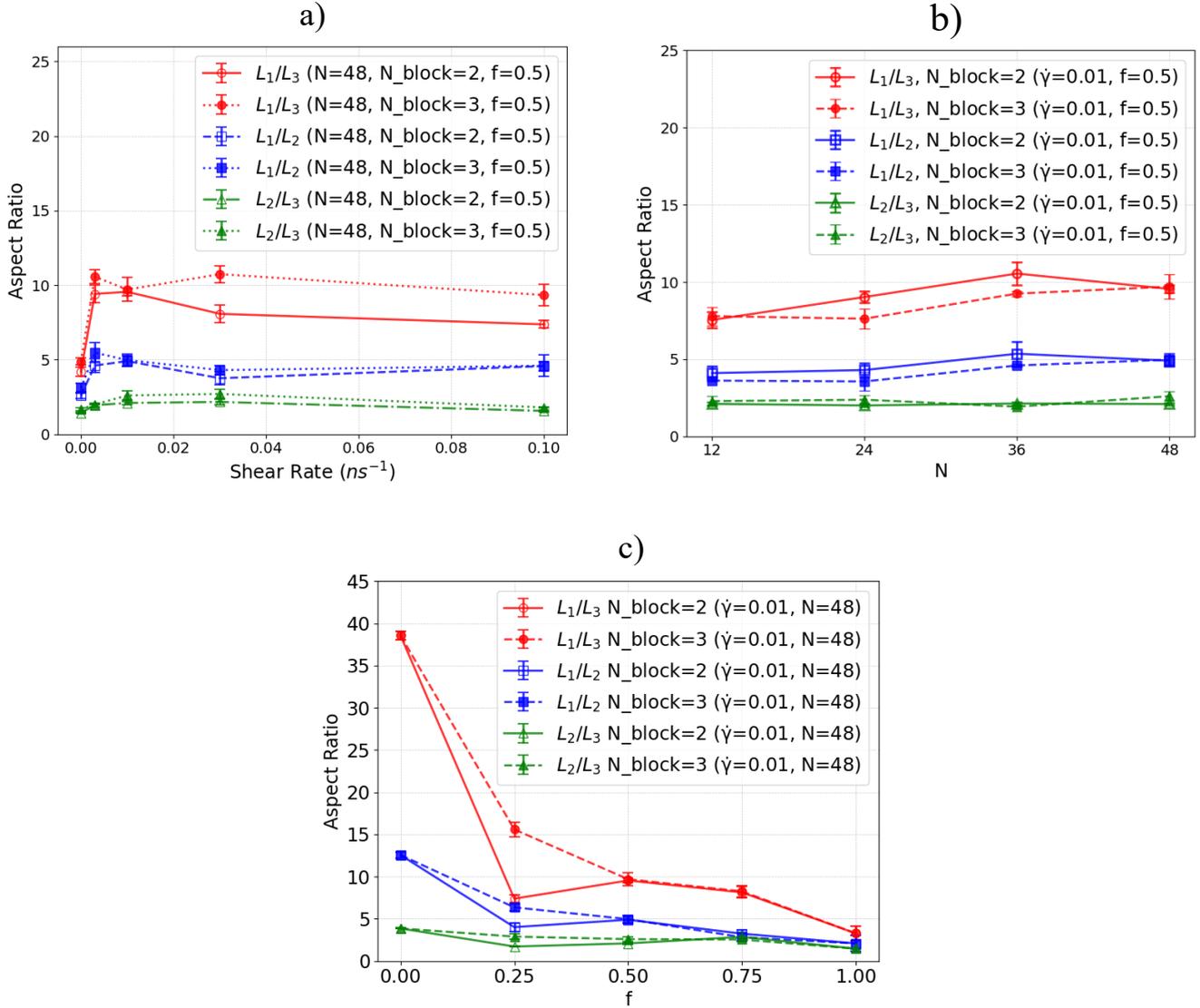

Figure 4: a) The impact of shear flow on the morphology of multi-chain diblock and triblock copolymers with 120 chains, chain length of N= 48 beads and composition parameter of 0.5; b) The effect of chain length on chain morphology of both multi-chain copolymers at f=0.5 and shear rate of 0.01 $ns^{-1}$; c) The impact of hydrophobic fraction on morphology of multi-chain block copolymer systems at N=48 and shear rate of 0.01 $ns^{-1}$

### 3.2. Gyration Radius and Cluster Analysis

In Figure 5, the gyration radius ($R_g$) serves as a probe of individual chain sizes, while cluster count analysis reveals the degree of multi-chain assemblies under varying conditions. The Figure 5a demonstrates a striking non-monotonic response to shear rate for $N_{chain}$=120 and N=48 beads. At equilibrium ($\dot{\gamma} = 0$), both diblock ($N_{block} = 2$) and triblock ($N_{block} = 3$) copolymers exhibit relatively compact conformations with $R_g \approx 5.5$ *nm*, reflecting the balance between intramolecular excluded volume and favourable hydrophobic associations. In Figure 5b, both architectures show dramatically elevated cluster counts at equilibrium (micelle count for diblocks ~8.5, micelle count for triblocks ~8), indicating many small, independent assemblies. The initial increase in Rg at low shear rates ($\dot{\gamma}$ = 0.003-0.01 $ns^{-1}$) represents an expected physical phenomenon: shear flow initially enhances chain extension and assembly aggregation before causing structural breakdown at higher rates. The triblock copolymers show a more dramatic increase, reaching Rg ≈ 12 *nm* at $\dot{\gamma}$ = 0.01 $ns^{-1}$, indicating their enhanced ability to form extended, interconnected networks through their bridging architecture. Diblock copolymers reach a peak Rg ≈ 9 *nm* at similar shear rates, reflecting their more limited extension capability. Simultaneously, cluster counts drop sharply for both architectures, with triblocks decreasing from 6.5 to ~3.8 and diblocks from 6.6 to ~4.8 at $\dot{\gamma}$ = 0.01 $ns^{-1}$, confirming that moderate shear promotes assembly aggregation into fewer, larger structures, as visualized in Figure 3a, which shows the network coalescing at $\dot{\gamma}$ = 0.003-0.01 $ns^{-1}$. The initial decrease in cluster count from equilibrium to $\dot{\gamma}$ = 0.01 $ns^{-1}$ represents shear-induced assembly aggregation, since shear flow provides activation energy that overcomes kinetic barriers. At higher shear rates ($\dot{\gamma}$ > 0.01 $ns^{-1}$), the Rg of diblock copolymers decreases monotonically to ~7.5 *nm* at $\dot{\gamma}$ = 0.1 $ns^{-1}$ as assemblies fragment and reorganize under excessive stress. Triblock copolymers maintain their elevated Rg ≈ 9 *nm* throughout the higher shear rate regime, demonstrating superior structural integrity and resistance to shear-induced breakdown. In Figure 5b, the cluster count behaviour diverges between architectures: diblocks show a slight increase from ~4.8 to ~5.5, indicating modest fragmentation, while triblocks maintain relatively stable counts around 4-4.5, confirming their mechanical stability under flow conditions (persistent interconnected structures in Figure 3a).

Figures 5c and 5d display the Rg and cluster count as a function of chain length, which show molecular size and assembly characteristics. Figure 5c shows that Rg increases systematically with chain length for both architectures at $\dot{\gamma}$ = 0.01 $ns^{-1}$ and f = 0.5. Diblock copolymers exhibit Rg values increasing from ~7 *nm* at N = 12 to ~9 *nm* at N = 48, consistent with the natural scaling of polymer dimensions. Triblock copolymers show a similar trend from ~8 *nm* to ~12 *nm*, maintaining consistently larger Rg values than diblocks at equivalent chain lengths, confirming their enhanced association propensity and ability to form more extended network structures. In Figure 5d, the cluster count analysis reveals systematic aggregation behaviour with increasing chain length. Both architectures exhibit lower cluster counts at short chain lengths (diblocks ~2, triblocks ~2.2 at N = 12), reflecting the formation of multiple small assemblies. As chain length increases, cluster counts rise to around 4 for both systems, showing an increasing trend for both diblocks and triblocks (large error bars are observed for triblock systems, with values approaching those of diblock systems). This behaviour indicates that longer chains enable the formation of more distinct assemblies, leading to network percolation visible in Figure 3b.

Figures 5e and 5f explore how hydrophobic fraction (f) affects individual chain size and cluster/micelle count. At f = 0.25, triblock chains exhibit higher Rg values (~8 *nm*) compared to diblocks (~5.5 *nm*) at $\dot{\gamma}$ = 0.01 $ns^{-1}$ with N = 48. In Figure 5e, from f = 0.25 to 0.75, both architectures show increasing trends, with triblocks consistently maintaining higher Rg values throughout. At f = 0.5, the Rg of triblock copolymers reaches ~12 *nm* while diblocks reach ~9 *nm*. Both peak at f = 0.75 with triblocks at ~13 *nm* and diblocks at ~12 *nm*. This parallel increasing behaviour indicates that growing hydrophobic content drives progressive chain stretching for both architectures. In terms of thermodynamic driving force, the energy gained through stronger hydrophobic interactions surpasses the energetic penalty associated with reduced chain entropy. For triblocks, higher Rg values reflect bridging configurations where chains span multiple assemblies. For diblocks, despite lacking central bridging blocks, individual chains still stretch as they participate in larger assemblies, though to a lesser extent. At f = 1.0, both architectures collapse to minimal Rg (~6 *nm*) due to phase separation without hydrophilic segments (single-domain structures in Figure 3c). The cluster count data show monotonic decreases from f = 0.25 to f = 1.0 for both architectures. Diblocks start higher (~7) and decrease to ~1, while triblocks maintain lower micelle/cluster counts throughout (~4 decreasing to ~1), reflecting more aggregation through bridging capability. The positive correlation between increasing Rg and decreasing cluster count confirms that chains stretch and assemblies aggregate, with triblocks showing enhanced extension through bridging.

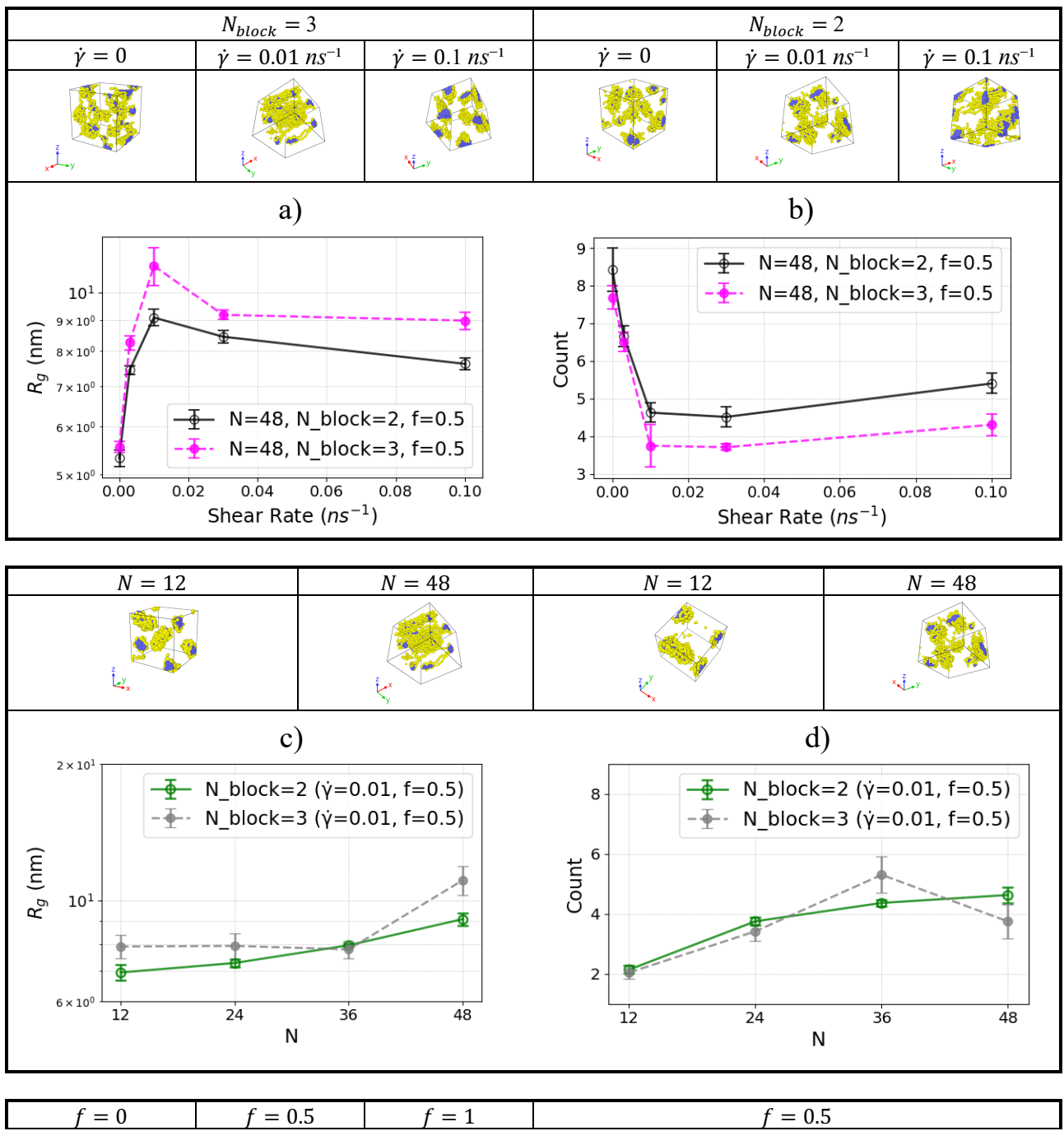

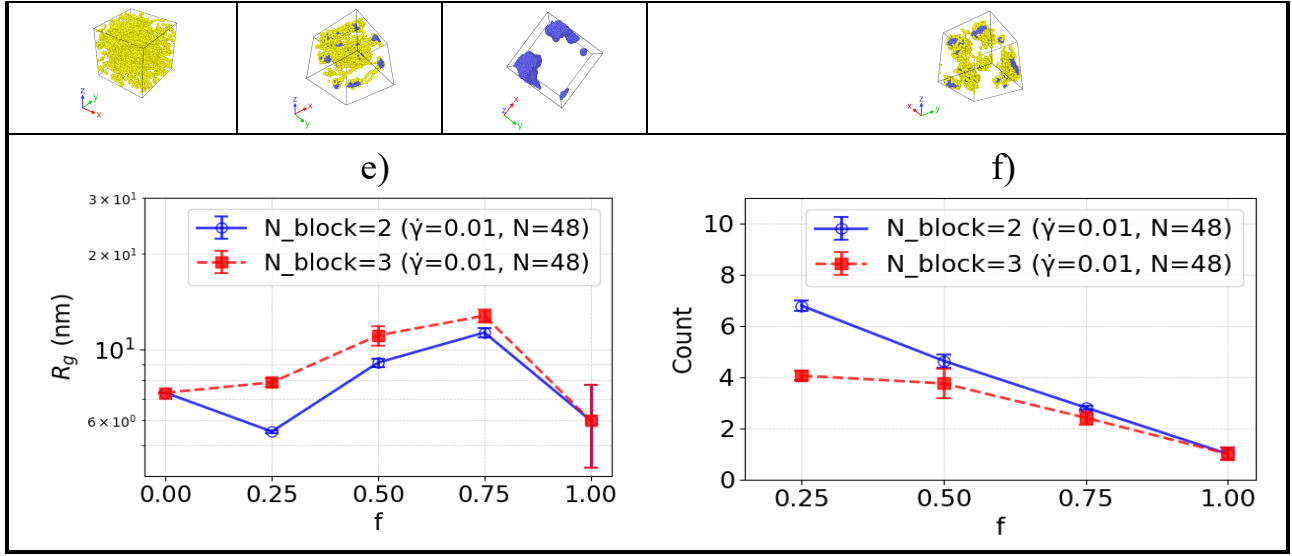

Figure 5: Gyration radius of individual chains and cluster count for multi-chain diblock and triblock copolymer systems with their snapshots at a and b) chain length of 48 beads and fraction parameter of 0.5; c and d) composition parameter of f=0.5 and shear rate of 0.01 $ns^{-1}$; e and f) shear rate of 0.01 $ns^{-1}$ and N=48

### 3.3. Solution Viscosity

In Figure 6, the solution viscosity ($\eta$) measurements provide crucial rheological insights into how block copolymer self-assembly under shear conditions translates into macroscopic flow properties, revealing the intimate connection between molecular architecture, assembled structure, and bulk material behaviour. Figure 6a demonstrates the strong shear-thinning behaviour for N=48 beads at f=0.5. At equilibrium conditions ($\dot{\gamma} = 0$), both diblock ($N_{block}$=2) and triblock ($N_{block}$=3) systems exhibit extremely high viscosities, with triblocks showing particularly elevated values around 12 $Pa.s$. This high zero-shear viscosity reflects the presence of large, interconnected assemblies that promote steric interactions between assemblies, and bridging networks that span the solution volume. The precipitous drop in viscosity with increasing shear rate represents classic shear-thinning behaviour [41], where the structured assemblies responsible for high zero-shear viscosity undergo progressive fragmentation and alignment. Triblock copolymers show a more dramatic decrease across the entire shear rate range, maintaining viscosities approximately half an order of magnitude higher than diblocks at weak shear rates, indicating that their more interconnected network structures create sustained flow resistance even under shear. This behaviour reflects the persistence of bridging networks that continue to provide mechanical reinforcement even at moderate shear rates. Viscosity trends correlate with structural evolution: from $\dot{\gamma} = 0$ to 0.01 $ns^{-1}$, viscosity decreases as cluster counts drop (8.5 to 4.8 for diblocks, 8 to 3.8 for triblocks, shown in Figure 5b) and networks coalesce (Figure 3a); at $\dot{\gamma} > 0.01$ $ns^{-1}$, viscosity decreases further as clusters increase (fragmentation) visible in snapshots in Figures 3a and 6d.

Figure 6b reveals how viscosity varies with chain length (N) at a shear rate of $\dot{\gamma} = 0.01$ $ns^{-1}$ for both architectures at f=0.5. Both systems show positive correlations with chain length, where diblocks ($N_{block}$=2) exhibit a modest increase in viscosity values, rising gradually from approximately 0.3 $Pa.s$ at N=12 to around 0.4 $Pa.s$ at N=48. Consistent with snapshots in Figure 3b, this gentle, nearly linear increase reflects the progressive growth of micellar assemblies with chain length, where larger discrete micelles contribute incrementally to higher solution viscosities through increased inter-micellar interactions. In contrast, triblocks ($N_{block}$=3) show a dramatic increase in viscosity with chain length, rising from approximately 0.35 *Pa.s* at N=12 to over 1.5 *Pa.s* at N=48. The increasing cluster count observed in snapshots in Figures 3b and 4b confirms the increase in viscosity with chain length for triblocks, since bridging networks begin to percolate through the solution, creating system-spanning structures that dramatically enhance resistance to flow. Triblock viscosity enhancement correlates with cluster count increase (Figure 5d: ~2.2 to ~4-5), and more clusters under constant volume fraction indicate larger percolating assemblies (Figure 3b: N=12 to N=48 progression). The architectural difference becomes increasingly pronounced at longer chain lengths, where bridging capability of the triblock can be fully exploited, while diblocks maintain their micellar structure, as manifested by their more modest viscosity enhancement.

Figure 6c depicts composition-dependent viscosity behaviour of diblock and triblock copolymer solutions. At f = 0, both architectures start at similar low viscosities (~0.15 *Pa.s*). From f = 0 to 0.5, triblocks show stronger increases (from ~0.15 *Pa.s* to ~1.5 *Pa.s*), while diblocks exhibit more modest growth (from ~0.15 *Pa.s* to ~0.4 *Pa.s*), indicating that triblock bridging configurations enhance network formation and inter-assembly connectivity more effectively. At f = 0.25-0.5, triblocks reach their maximum viscosity, reflecting optimal bridging network structures, where hydrophobic associations strengthen connectivity, while sufficient hydrophilic segments maintain solvation. Beyond f = 0.5, triblock viscosity decreases to ~0.9 at f = 0.75, as increasing hydrophobic content reduces hydrophilic segment length, weakening network connectivity. Diblocks show distinctly different behaviour: a gradual increase from f = 0 to f = 1.0, rising from ~0.15 *Pa.s* to ~1 *Pa.s*. Without central bridging blocks, diblocks form discrete micelles rather than extended networks, with viscosity increasing steadily as hydrophobic content grows and micelle structures aggregate, leading to a decrease in micelle count (shown in snapshots in Figure 3c and Figure 5f). At f = 1.0, both architectures converge to similar values (~1 *Pa.s*), where complete phase separation dominates and architectural differences become insignificant. The contrasting trends highlight different underlying mechanisms leading to a viscosity increase, where triblocks achieve maximum viscosity at intermediate compositions through bridging-based network connectivity, while diblocks show progressive increases driven by discrete micelle growth.

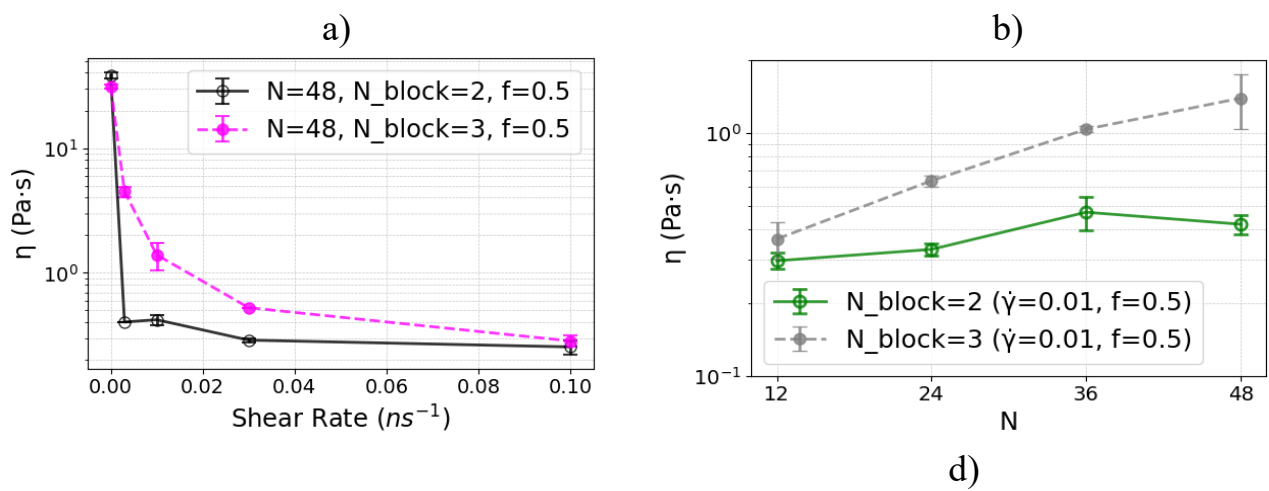

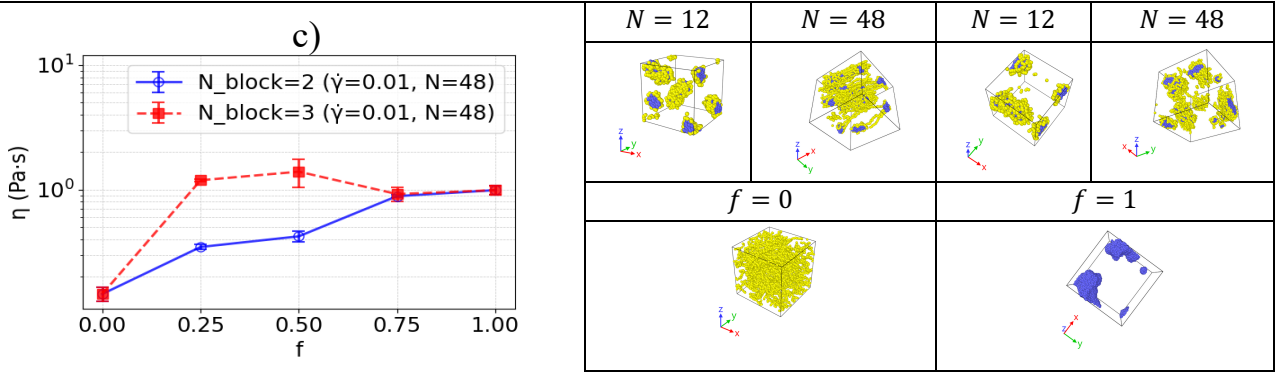

Figure 6: Solution viscosity ($\eta$) of both diblock and triblock copolymers at a) various shear rates, fixed chain length, N=48, and composition parameter, f=0.5; b) different chain lengths, fixed shear rate of 0.01 $ns^{-1}$ and f=0.5; c) various f, fixed N=48, and shear rate of 0.01 $ns^{-1}$; and d) the snapshots of block copolymers at different shear rates, chain lengths, and hydrophobic fractions

### 3.4. Viscoelasticity and Terminal Relaxation Time

In Figures 7a and 7b, at N = 48 and fraction parameter f = 0.5, the storage modulus (G′) increases with frequency while the loss modulus (G″) dominates at low frequencies, indicating viscous flow behaviour at long timescales [38]. The storage (G′) and loss (G″) moduli are obtained from the stress relaxation modulus, G(t), whose calculation is detailed in the Supplementary Materials [Section S1]. Two well-separated relaxation processes are expected in micellar solutions: a fast process governing redistribution of micellar sizes through single-chain insertion and expulsion events at constant total micelle number [43], and a slow process governing collective reorganisation of the micellar population through formation and dissolution events [44]. These two processes give rise to two distinct crossover frequencies in the viscoelastic spectra, where G′ = G″. The higher-frequency crossover reflects rapid single-chain exchange dynamics, while the lower-frequency crossover reflects the much slower collective structural reorganisation of the micellar population. The terminal relaxation time extracted as $1/\omega_c$ from the primary lower-frequency crossover in Figure 7 therefore characterises the timescale over which the micellar solution fully relaxes its structure through collective formation and dissolution events.

The two crossover features observed at short chain lengths (N = 12 and N = 24) in Figures 7c and 7d directly confirm the coexistence of both relaxation processes within the accessible frequency window. As chain length increases toward N = 36 and N = 48, the energetic barrier governing single-chain expulsion grows with increasing core block size, causing the fast process to accelerate beyond the accessible frequency range and leaving only the slower collective reorganisation process visible as a single crossover at lower frequency. This progression shows that the two relaxation timescales become increasingly well-separated as chain length grows, consistent with the structural evolution observed in Figures 3b and 5c, showing progressive formation of larger, more stable assemblies. Comparing diblock and triblock systems at identical conditions, triblock copolymers exhibit higher storage moduli at high frequencies, reflecting the additional elastic constraints imposed by bridging configurations, where the central hydrophilic block connects distinct micellar aggregates into a transient network. This enhanced elasticity is evidenced by the lower cluster counts in Figure 5b and the more interconnected structures visible in Figure 3a. The high-frequency plateau region reflects the capacity of this transient network to store elastic energy before individual chains can exchange between aggregates. In Figures 7e and 7f, at N = 48 and $\dot{\gamma}$ = 0.01 $ns^{-1}$, the effect of hydrophobic fraction on the viscoelastic response reveals fundamentally different behaviours between the two architectures. At f = 0, both architectures exhibit purely viscous behaviour with G″ dominating across the high frequency range, consistent with unassociated polymer solutions lacking any micellar structure. As f increases, the growing hydrophobic content drives progressive micellisation and network formation, increasing both G′ and G″. At f = 1.0, both architectures form densely packed hydrophobic domains with enhanced steric constraints that increase elastic response, raising G′ relative to the zero-composition case.

In Figure 8a, diblock copolymers exhibit an overall decreasing terminal relaxation time with increasing shear rate. As shear progressively disrupts and fragments discrete diblock micelles, as evidenced by the increasing cluster counts at $\dot{\gamma} > 0.01$ $ns^{-1}$ in Figure 5b, smaller and more numerous assemblies form with reduced structural integrity. These smaller assemblies undergo faster collective reorganisation, reducing the terminal relaxation time overall. Triblock copolymers show a monotonic increase in terminal relaxation time with shear rate because the progressive loss of bridging network connectivity under shear, visible in Figure 3a, eliminates fast parallel relaxation pathways. The remaining network must relax through increasingly constrained and cooperative reorganisation events involving fewer but more strongly interconnected junctions, requiring progressively longer timescales as shear rate increases. In Figure 8b, the terminal relaxation time decreases with increasing chain length N for both architectures at $\dot{\gamma} = 0.01$ $ns^{-1}$, with triblocks maintaining slightly higher values than diblocks. As chain length increases, longer chains accumulate greater drag forces that are transmitted through the polymer backbone to the anchored hydrophobic blocks, effectively lowering the energetic barrier for junction disruption and accelerating collective micellar reorganisation, reducing the terminal relaxation time. The slightly higher values maintained by triblocks reflect the double-ended bridging architecture, where complete chain exchange requires both hydrophobic end blocks to sequentially escape from their respective cores, inherently slowing the collective relaxation relative to single-ended diblock exchange.

In Figure 8c, triblock systems show increasing terminal relaxation time with f, as stronger hydrophobic associations and growing hydrophobic block size progressively deepen the energetic barrier for chain expulsion, while double-ended bridging configurations create increasingly cooperative network junctions that require collective rather than independent dissolution events. The simultaneous decrease in cluster counts from ~4 to ~2.5 (Figure 5f) confirms that micelles merge into fewer but larger and more stable assemblies whose collective reorganisation requires progressively longer timescales. Diblock systems maintain nearly constant terminal relaxation time (~1140–1150 $ns$) across all compositions because single-ended hydrophobic associations form discrete, composition-insensitive micellar structures whose collective reorganisation dynamics are not significantly altered by the growing hydrophobic content within the accessible composition range. These contrasting trends demonstrate that polymer architecture fundamentally governs the collective micellar relaxation dynamics through distinct physical mechanisms: composition-independent discrete micellar reorganisation for diblocks versus composition-sensitive cooperative network dissolution driven by double-ended bridging for triblocks.

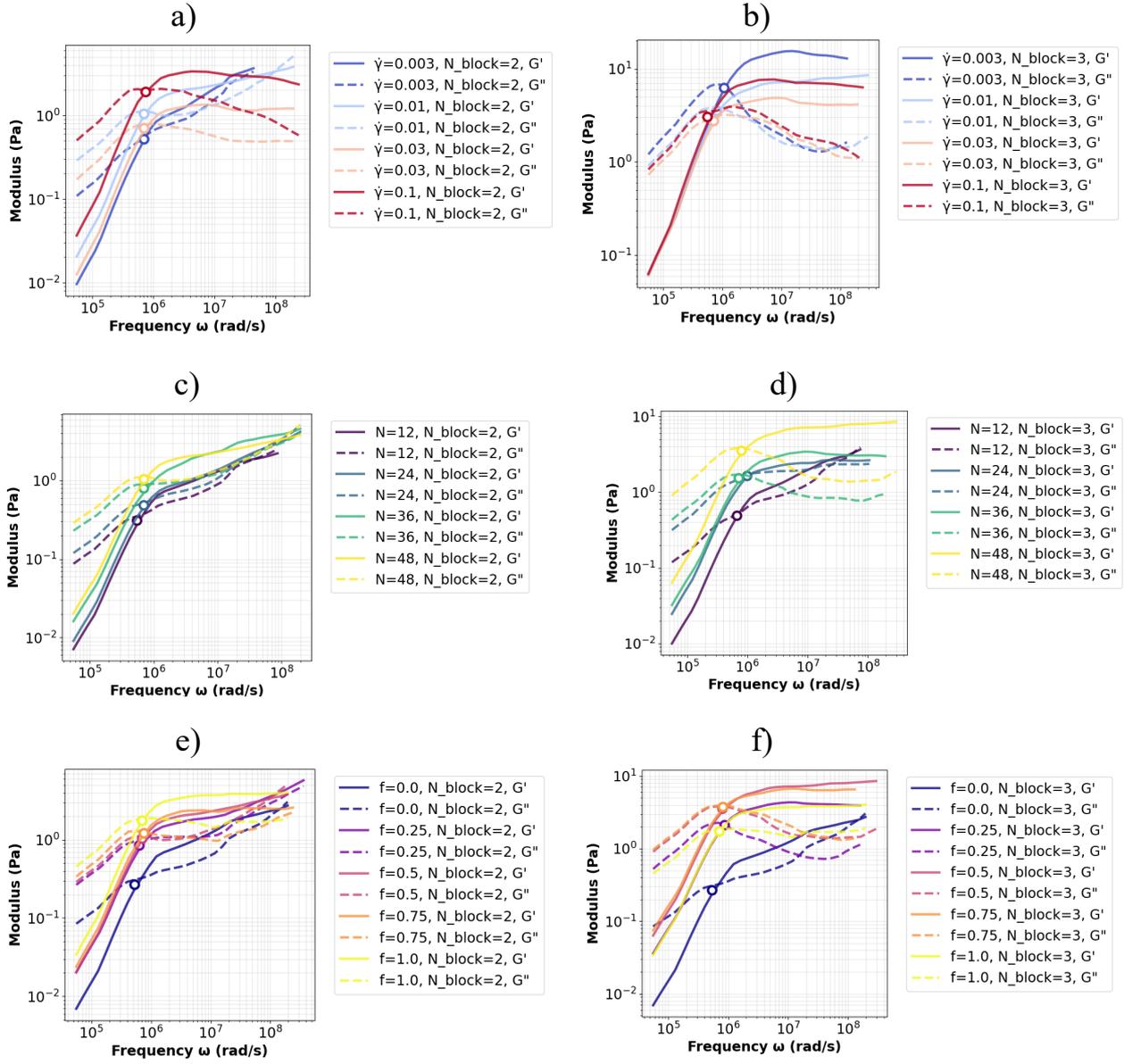

Figure 7: The viscoelastic (*G′* and *G″*) behaviour of both diblock and triblock copolymers as a function of frequency at a and b) various shear rates, fixed N=48 and fraction parameter of 0.5; c and d) different chain lengths, shear rate of 0.01 $ns^{-1}$, and fraction parameter of 0.5; e and f) various hydrophobic fractions, N=48 and shear rate of 0.01 $ns^{-1}$

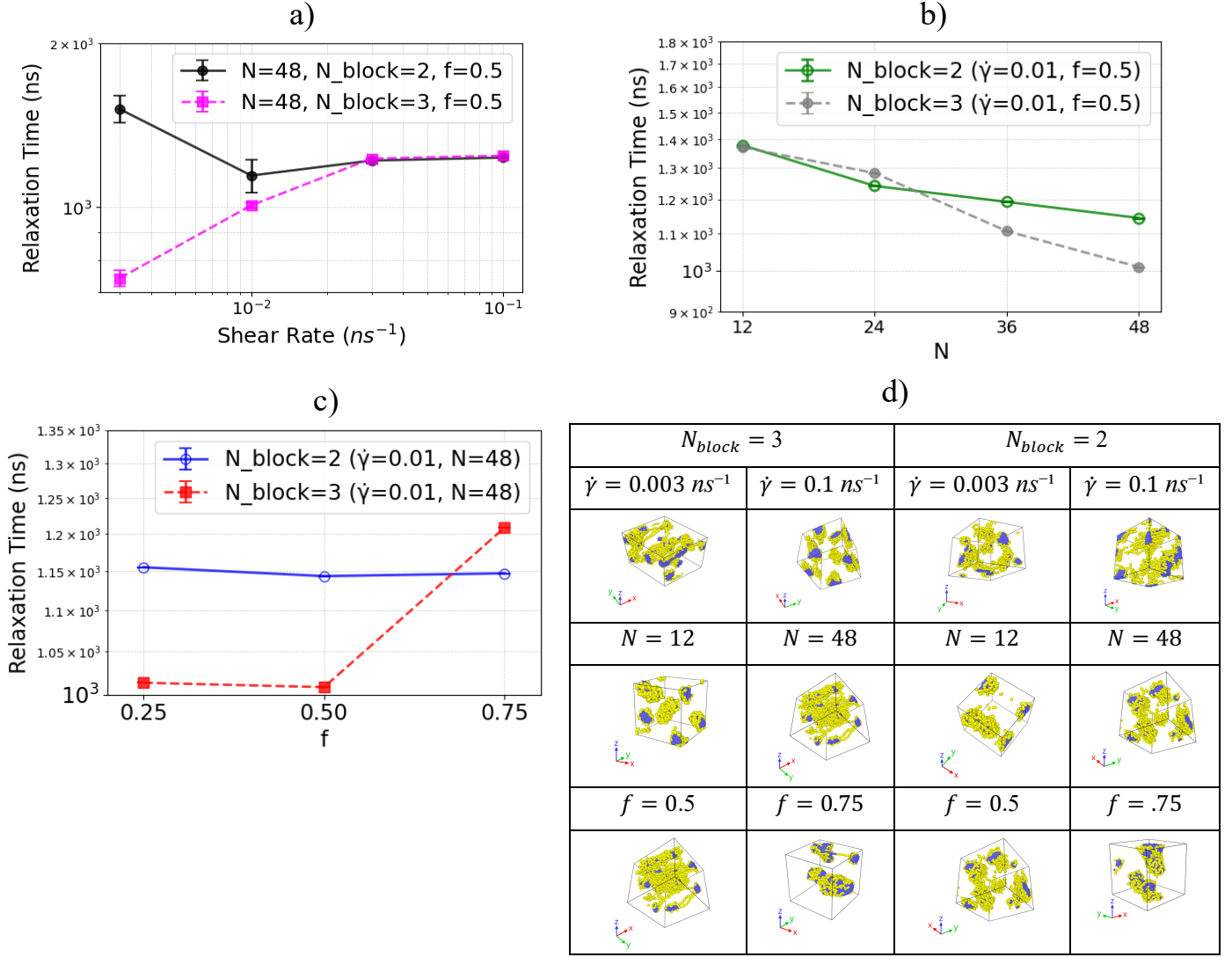

Figure 8: Terminal relaxation time of diblock and triblock copolymers at a) different shear rates, N=48 and f=0.5; b) various chain lengths, shear rate of 0.01 $ns^{-1}$ and fixed f=0.5; c); different fraction parameters, N=48 and shear rate of 0.01 $ns^{-1}$; and d) the snapshots of block copolymers at various shear rates, chain lengths, and chemical compositions

## 4. Conclusion

This comprehensive computational study establishes the fundamental mechanisms by which multi-chain amphiphilic block copolymer systems in dilute solutions are assembled. We have investigated how molecular architecture, chain length, composition, and flow conditions collectively govern the self-assembly and rheological behaviour of these systems. Using Brownian dynamics simulations validated against established polymer solution scaling theory, we have uncovered the molecular mechanisms through which architectural choices translate into macroscopic material performance. The main findings of this research are summarized as follows:

- Bridging-capable triblock architectures generate interconnected network structures that produce viscosities half an order of magnitude greater than diblock systems, with enhanced mechanical resistance under weak flow conditions.

- Both architectures exhibit non-monotonic responses to applied flow, with initial assembly aggregation at weak shear ($\dot{\gamma}$ = 0.003-0.01 $ns^{-1}$), transitioning to fragmentation at high rates.

- Triblocks achieve extreme prolate deformations (aspect ratios ~11) through network elongation, while diblocks form less anisotropic discrete micelles (aspect ratios ~7.5), reflecting fundamentally different assembly mechanisms.

- Four distinct morphologies of multi-chain block copolymers were identified, including spherical micelles, cigar-shaped micelles, short cylinder-like micelles, and worm-like micelles.

- Increasing chain length drives systematic structural evolution, with triblocks displaying dramatic viscosity enhancement, suggesting percolation transitions, whereas diblocks show modest linear increases consistent with micellar growth.

- Frequency-dependent moduli reveal that triblocks exhibit progressively slower relaxation with increasing hydrophobic content, while diblocks maintain relatively constant dynamics.

- Terminal relaxation time measurements directly connect macroscopic rheology to junction stability, with triblocks showing composition-sensitive lifetimes (1000-1200 $ns$), reflecting cooperative double-ended associations, versus composition-independent diblock behaviour (~1140-1150 $ns$).

## Supporting information

The Supporting Information is available here.

Figure S1: Simulation procedure for multi-chain block copolymers includes eight stages; step 1) defining parameter space using Signac; step 2) generating input files and initial configurations depending on input parameters; step 3) implementing energy minimization and annealing process preventing system from kinetic trapping; step 4) applying BD simulation in LAMMPS package under shear and non-shear flows to capture conservative, random and friction forces; step 5) real-time monitoring during t=2200 $ns$; step 6) post-processing analysis for extracting target parameters from LAMMPS and Python-script code; step 7) morphology-dependent analysis based on hydrophobic fraction, shear rate, chain length, chain number, and molecular architecture using required Python libraries; step 8) doing statistical analysis when extracting final target parameters. Figure S2: Comparison of radius of gyration of homopolymers with chain length of *N*=16, 32, 48, 64, and 80 with Flory theory at equilibrium under a) good; and b) poor solvent conditions.

## Conflicts of interest

There is no conflict of interest to be declared.

## Acknowledgements

The first author gratefully acknowledges support from the Melbourne Research Scholarship (MRS) awarded by the University of Melbourne. This work was also supported by the University of Melbourne's Research Computing Services.

# References

[1]  D. A. Z. Wever, F. Picchioni, and A. A. Broekhuis, "Polymers for enhanced oil recovery: A paradigm for structure-property relationship in aqueous solution," *Prog. Polym. Sci.*, vol. 36, no. 11, pp. 1558–1628, 2011, doi: 10.1016/j.progpolymsci.2011.05.006.

[2]  F. A. Chapa-Villarreal, M. Miller, J. J. Rodriguez-Cruz, D. Pérez-Carlos, and N. A. Peppas, "Self-assembled block copolymer biomaterials for oral delivery of protein therapeutics," *Biomaterials*, vol. 300, no. May, 2023, doi: 10.1016/j.biomaterials.2023.122191.

[3]  H. Chu and C. C. Wang, "Metal-organic frameworks meet synthetic polymers for water decontamination: A critical review," *Chem. Eng. J.*, vol. 476, no. September, p. 146684, 2023, doi: 10.1016/j.cej.2023.146684.

[4]  P. Samaddar, A. Deep, and K. H. Kim, "An engineering insight into block copolymer self-assembly: Contemporary application from biomedical research to nanotechnology," *Chem. Eng. J.*, vol. 342, no. October 2017, pp. 71–89, 2018, doi: 10.1016/j.cej.2018.01.062.

[5]  E. Brodszkij *et al.*, "Poly(Sitosterol)-Based Hydrophobic Blocks in Amphiphilic Block Copolymers for the Assembly of Hybrid Vesicles," *Small*, vol. 20, no. 40, pp. 1–17, 2024, doi: 10.1002/smll.202401934.

[6]  Z. Li and Z. Lin, "Self-assembly of block copolymers for biological applications," *Polym. Int.*, vol. 71, no. 4, pp. 366–370, 2022, doi: 10.1002/pi.6327.

[7]  A. O. Moughton, M. A. Hillmyer, and T. P. Lodge, "Multicompartment block polymer micelles," *Macromolecules*, vol. 45, no. 1, pp. 2–19, 2012, doi: 10.1021/ma201865s.

[8]  Y. Y. Won, A. K. Brannan, H. T. Davis, and F. S. Bates, "Cryogenic transmission electron microscopy (cryo-TEM) of micelles and vesicles formed in water by poly(ethylene oxide)-based block copolymers," *J. Phys. Chem. B*, vol. 106, no. 13, pp. 3354–3364, 2002, doi: 10.1021/jp013639d.

[9]  T. M. Birshtein and E. B. Zhulina, "Scaling theory of supermolecular structures in block copolymer-solvent systems: 2. Supercrystalline structures," *Polymer (Guildf).*, vol. 31, no. 7, pp. 1312–1320, 1990, doi: 10.1016/0032-3861(90)90223-L.

[10] E. B. Zhulina and O. V. Borisov, "Theory of block polymer micelles: Recent advances and current challenges," *Macromolecules*, vol. 45, no. 11, pp. 4429–4440, 2012, doi: 10.1021/ma300195n.

[11] G. Campos-Villalobos, F. R. Siperstein, A. Charles, and A. Patti, "Solvent-induced morphological transitions in methacrylate-based block-copolymer aggregates," *J. Colloid Interface Sci.*, vol. 572, pp. 133–140, 2020, doi: 10.1016/j.jcis.2020.03.067.

[12] J. B. Sabadini, C. L. P. Oliveira, and W. Loh, "Do ethoxylated polymeric coacervate micelles respond to temperature similarly to ethoxylated surfactant aggregates?," *J. Colloid Interface Sci.*, vol. 678, no. PA, pp. 1012–1021, 2025, doi: 10.1016/j.jcis.2024.08.248.

[13] S. Burke and A. Eisenberg, "Physico-chemical investigation of multiple asymmetric amphiphilic diblock copolymer morphologies in solution," *High Perform. Polym.*, vol. 12, no. 4, pp. 535–542, 2000, doi: 10.1088/0954-0083/12/4/308.

[14] N. Moreno, S. Nunes, and V. Calo, "Morphological Transitions of Block Copolymer Micelles: Implications for Mesoporous Materials Ordering," *Macromol. Theory Simulations*, vol. 34, no. 1, pp. 1–13, 2025, doi: 10.1002/mats.202400046.

[15] M. J. Hafezi and F. Sharif, "Brownian dynamics simulation of amphiphilic block copolymers with different tail lengths, comparison with theory and comicelles," *J. Mol. Graph. Model.*, vol. 62, pp. 165–173, 2015, doi: 10.1016/j.jmgm.2015.09.005.

[16] M. J. Cass, D. M. Heyes, and R. J. English, "Brownian dynamics simulations of associating diblock copolymers," *Langmuir*, vol. 23, no. 12, pp. 6576–6587, 2007, doi: 10.1021/la063210j.

[17] J. Zhang, Z. Y. Lu, and Z. Y. Sun, "Self-assembly structures of amphiphilic multiblock copolymer in dilute solution," *Soft Matter*, vol. 9, no. 6, pp. 1947–1954, 2013, doi: 10.1039/c2sm27092g.

[18] P. Hao, X. H. Mai, Q. Y. Chen, and M. M. Ding, "Conformation of an Amphiphilic Comb-like Copolymer in a Selective Solvent," *Chinese J. Polym. Sci. (English Ed.*, vol. 41, no. 9, pp. 1386–1391, 2023, doi: 10.1007/s10118-023-2912-8.

[19] S. Liu and R. Sureshkumar, "Morphological Diversity in Diblock Copolymer Solutions: A Molecular Dynamics Study," *Colloids and Interfaces*, vol. 7, no. 2, 2023, doi: 10.3390/colloids7020040.

[20] S. Liu, M. S. Samie, and R. Sureshkumar, "Vesicle Morphogenesis in Amphiphilic Triblock Copolymer Solutions," *Colloids and Interfaces*, vol. 8, no. 3, 2024, doi: 10.3390/colloids8030029.

[21] G. Srinivas, D. E. Discher, and M. L. Klein, "Self-assembly and properties of diblock copolymers by coarse-grain molecular dynamics," *Nat. Mater.*, vol. 3, no. 9, pp. 638–644, 2004, doi: 10.1038/nmat1185.

[22] D. Bedrov, C. Ayyagari, and G. D. Smith, "Multiscale modeling of poly(ethylene oxide)-poly(propylene oxide)-poly(ethylene oxide) triblock copolymer micelles in aqueous solution," *J. Chem. Theory Comput.*, vol. 2, no. 3, pp. 598–606, 2006, doi: 10.1021/ct050334k.

[23] T. T. Pham, B. Dünweg, and J. R. Prakash, "Collapse dynamics of copolymers in a poor solvent: Influence of hydrodynamic interactions and chain sequence," *Macromolecules*, vol. 43, no. 23, pp. 10084–10095, 2010, doi: 10.1021/ma101806n.

[24] A. Q. Liu, L. J. Liu, W. S. Xu, X. L. Xu, J. Z. Chen, and L. J. An, "Stress-Structure Relationship of the Reversible Associating Polymer Network under Start-up Shear Flow," *Chinese J. Polym. Sci. (English Ed.*, vol. 39, no. 3, pp. 387–396, 2021, doi: 10.1007/s10118-020-2487-6.

[25] D. Amin and Z. Wang, "Nonlinear rheology and dynamics of supramolecular polymer networks formed by associative telechelic chains under shear and extensional flows," *J. Rheol. (N. Y. N. Y).*, vol. 64, no. 3, pp. 581–600, 2020, doi: 10.1122/1.5120897.

[26] N. Sugimura and K. Ohno, "A Monte Carlo simulation of water + oil + ABA triblock copolymer ternary system II. Rheology under shear flow field by Monte Carlo Brownian Dynamics method," *Chem. Phys. Lett.*, vol. 777, no. January, p. 138382, 2021, doi: 10.1016/j.cplett.2021.138382.

[27] A. Nikoubashman, R. A. Register, and A. Z. Panagiotopoulos, "Simulations of shear-

induced morphological transitions in block copolymers," *Soft Matter*, vol. 9, no. 42, pp. 9960–9971, 2013, doi: 10.1039/c3sm51759d.

[28] P. Xu, J. Lin, L. Wang, and L. Zhang, "Shear flow behaviors of rod-coil diblock copolymers in solution: A nonequilibrium dissipative particle dynamics simulation," *J. Chem. Phys.*, vol. 146, no. 18, 2017, doi: 10.1063/1.4982938.

[29] L. Schneider and M. Müller, "Rheology of symmetric diblock copolymers," *Comput. Mater. Sci.*, vol. 169, no. June, p. 109107, 2019, doi: 10.1016/j.commatsci.2019.109107.

[30] S. J. Marrink, H. J. Risselada, S. Yefimov, D. P. Tieleman, and A. H. De Vries, "The MARTINI force field: Coarse grained model for biomolecular simulations," *J. Phys. Chem. B*, vol. 111, no. 27, pp. 7812–7824, 2007, doi: 10.1021/jp071097f.

[31] F. Mu, D. Reith, and M. Pu, "from Atomistic Simulations," *J. Comput. Chem.*, 2003.

[32] Z. Shireen *et al.*, "A machine learning enabled hybrid optimization framework for efficient coarse-graining of a model polymer," *npj Comput. Mater.*, vol. 8, no. 1, pp. 1–11, 2022, doi: 10.1038/s41524-022-00914-4.

[33] H. Weeratunge *et al.*, "Bayesian coarsening: rapid tuning of polymer model parameters," *Rheol. Acta*, vol. 62, no. 10, pp. 477–490, 2023, doi: 10.1007/s00397-023-01397-w.

[34] R. Everaers, H. A. Karimi-Varzaneh, F. Fleck, N. Hojdis, and C. Svaneborg, "Kremer-Grest Models for Commodity Polymer Melts: Linking Theory, Experiment, and Simulation at the Kuhn Scale," *Macromolecules*, vol. 53, no. 6, pp. 1901–1916, 2020, doi: 10.1021/acs.macromol.9b02428.

[35] K. H. Shen, M. Fan, and L. M. Hall, "Molecular Dynamics Simulations of Ion-Containing Polymers Using Generic Coarse-Grained Models," *Macromolecules*, vol. 54, no. 5, pp. 2031–2052, 2021, doi: 10.1021/acs.macromol.0c02557.

[36] N. Clisby, "Accurate estimate of the critical exponent ν for self-avoiding walks via a fast implementation of the pivot algorithm," *Phys. Rev. Lett.*, vol. 104, no. 5, pp. 1–4, 2010, doi: 10.1103/PhysRevLett.104.055702.

[37] Flory, P. J. "Principles of Polymer Chemistry", *Cornell University Press*, 1953.

[38] A. Tavera-Vázquez, B. Arenas-Gómez, C. Garza, Y. Liu, and R. Castillo, "Structure, rheology, and microrheology of wormlike micelles made of PB-PEO diblock copolymers," *Soft Matter*, vol. 14, no. 35, pp. 7264–7276, 2018, doi: 10.1039/c8sm01530a.

[39] H. Watanabe, M. L. Yao, T. Sato, and K. Osaki, "Non-newtonian flow behavior of diblock copolymer micelles: Shear-thinning in a nonentangling matrix," *Macromolecules*, vol. 30, no. 19, pp. 5905–5912, 1997, doi: 10.1021/ma961867d.

[40] M. Nguyen-Misra, S. Misra, Y. Wang, K. Rodrigues, and W. L. Mattice, "Simulation of self-assembly in solution by triblock copolymers with sticky blocks at their ends," *Prog. Colloid Polym. Sci.*, vol. 103, pp. 138–145, 1997, doi: 10.1007/3-798-51084-9_16.

[41] A. Prhashanna, S. A. Khan, and S. B. Chen, "Micelle morphology and chain conformation of triblock copolymers under shear: LA-DPD study," *Colloids Surfaces A Physicochem. Eng. Asp.*, vol. 506, pp. 457–466, 2016, doi: 10.1016/j.colsurfa.2016.07.003.

[42] S. Liu and R. Sureshkumar, "Deformation, Rupture, and Morphology Hysteresis of

Copolymer Nanovesicles in Uniform Shear Flow," *Langmuir*, vol. 41, no. 8, pp. 5083–5096, 2025, doi: 10.1021/acs.langmuir.4c04200.

[43]  E. E. Dormidontova, "Micellization kinetics in block copolymer solutions: scaling model," *Macromolecules*, vol. 32, no. 22, pp. 7630–7644, 1999, doi: 10.1021/ma9809029.

[44]  A. Halperin and S. Alexander, "Polymeric Micelles: Their Relaxation Kinetics," *Macromolecules*, vol. 22, no. 5, pp. 2403–2412, 1989, doi: 10.1021/ma00195a069.

[45]  E. Hajizadeh and R. G. Larson, " Stress-gradient-induced polymer migration in Taylor–Couette flow", *Soft Matter*, vol. 13, no. 35, pp. 5942-5949, 2017, DOI https://doi.org/10.1039/C7SM00821J

[46]  E. Hajizadeh and H. Garmabi, " Response Surface Based Optimization of Toughness of Hybrid Polyamide 6 Nanocomposites", *International Journal of Materials and Metallurgical Engineering*, vol. 1, no. 10, 2007.

[47]  J Xiang, E Hajizadeh, RG Larson, D Nelson, "Predictions of polymer migration in a dilute solution between rotating eccentric cylinders", *Journal of Rheology*, vol. 65, no. 6, pp. 1311-1325, 2021. https://doi.org/10.1122/8.0000330.

[48]  J Panisilvam, E Hajizadeh, H Weeratunge, J Bailey, S Kim, " Asymmetric cyclegans for inverse design of photonic metastructures", *APL Machine Learning*, vol.1, no.4, pp. 046105, 2023, https://doi.org/10.1063/5.0159264.

[49]  H. Weeratunge, D. M. Robe, E. Hajizadeh, " Interpretable SHAP-bounded Bayesian optimization for underwater acoustic metamaterial coating design", *Structural and Multidisciplinary Optimization*, vol. 68, no. 9, pp. 175, 2025, https://doi.org/10.1007/s00158-025-04104-w.

[50]  J Abdolahi, D Robe, RG Larson, M Kirley, E Hajizadeh, " Interpretable active learning meta-modeling for the association dynamics of telechelic polymers on colloidal particles", Journal of Rheology, vol. 69, no. 2, pp. 183-199, 2025, https://doi.org/10.1122/8.0000930.

[51]  D. M. Robe, A. Menzel, A. W. Phillips, E. Hajizadeh, " From SMILES to scattering: Automated high-throughput atomistic polyurethane simulations compared with WAXS data", *Computational Materials Science*, vol. 256, pp. 113931, 2025, https://doi.org/10.1016/j.commatsci.2025.113931

[52]  P. Ding, D. Robe, M. Kirley, E. Hajizadeh, " Constrained Bayesian accelerated design of acoustic polyurethane coatings with metamaterial features under hydrostatic pressure", *Structural and Multidisciplinary Optimization*, vol. 68, no. 8, pp. 159, 2025, https://doi.org/10.1007/s00158-025-04107-7

[53]  H. Weeratunge, Z. Shireen, S. Iyer, A. Menzel, A. W. Phillips, S. Halgamuge, R. Sandberg, E. Hajizadeh, "A machine learning accelerated inverse design of underwater acoustic polyurethane coatings", Structural and Multidisciplinary Optimization 65, 213 (2022). https://doi.org/10.1007/s00158-022-03322-w

[54]  Z. Shireen, H. Weeratunge, A. Menzel, A. w. Phillips, R. G Larson, K. Smith-Miles, E. Hajizadeh, "A machine learning enabled hybrid optimization framework for efficient coarse-graining of a model polymer", n*pj Computational Materials, vol.* 8, pp. 224 (2022). https://doi.org/10.1038/s41524-022-00914-4

[55] H. Weeratunge, D. Robe, A. Menzel, A. W. Phillips, M. Kirley, K. Smith-Miles, E. Hajizadeh, "Bayesian coarsening: rapid tuning of polymer model parameters", *Rheologica Acta*, vol. 62, pp. 477-490, 2023. https://doi.org/10.1007/s00397-023-01397-w

[56] A Jayawardena, A Hung, G Qiao, E Hajizadeh, "Molecular dynamics simulations of structurally nanoengineered antimicrobial peptide polymers interacting with bacterial cell membranes", *The Journal of Physical Chemistry B*, vol. 129, no. 1, pp. 250-259, 2024. https://doi.org/10.1021/acs.jpcb.4c06691.

[57] A Jayawardena, A Hung, G Qiao, E Hajizadeh, "Molecular Dynamics Simulation of the Interaction of Lipidated Structurally Nano Engineered Antimicrobial Peptide Polymers with Bacterial Cell Membrane", *The Journal of Physical Chemistry B, vol.* 129, no. 37, pp. 9382-9393, 2025, https://doi.org/10.1021/acs.jpcb.5c02067.

[58] A Jayawardena, A Hung, G Qiao, N O'Brien-Simpson, E Hajizadeh, "Membrane Selectivity of Star-Shaped Peptides: A Comparative Molecular Dynamics Study across Different Bacterial and Mammalian Bilayers", *The Journal of Physical Chemistry B, vol.* 129, no. 46, pp. 11983-11994, 2025. https://doi.org/10.1021/acs.jpcb.5c06308.

[59] G Zhu, H Rezvantalab, E Hajizadeh, X Wang, RG Larson, " Stress-gradient-induced polymer migration: Perturbation theory and comparisons to stochastic simulations", *Journal of Rheology*, vol. 60, no. 2, pp. 327-343, 2016. https://doi.org/10.1122/1.4942252

[60] E. Hajizadeh, B. D. Todd, P. J. Daivis, "A molecular dynamics investigation of the planar elongational rheology of chemically identical dendrimer-linear polymer blends", The Journal of Chemical Physics, vol. 142, no. 17, 2015. https://doi.org/10.1063/1.4919654.